\documentclass[12pt]{iopart}

\usepackage{iopams}

\usepackage{amsthm}

\usepackage{mathrsfs}
\usepackage{amssymb}

\newtheorem{theorem}{Theorem}

\newtheorem{prop}{Proposition}

\def\be{\begin{equation}}
\def\ee{\end{equation}}
\def\bea{\begin{eqnarray}}
\def\eea{\end{eqnarray}}
\def\bean{\begin{eqnarray*}}
\def\eean{\end{eqnarray*}}

\def\fin{\hfill \rule{2.5mm}{2.5mm}\\ \vspace{0mm}}
\def\finn{\hfill \rule{2.5mm}{2.5mm}}

\def\espaitemps{({\cal V},g)}
\def\varietat{{\cal V}}

\newcommand\ben{\begin{enumerate}}
\newcommand\een{\end{enumerate}}
\newcommand\bit{\begin{itemize}}
\newcommand\eit{\end{itemize}}

\newcommand\g{\gamma}
\newcommand\s{\sigma}
\newcounter{mnotecount}

\begin{document}

\rapid[Singularity theorems]{Singularity theorems based on trapped submanifolds of arbitrary co-dimension}

\author{Gregory J Galloway$^1$ and Jos\'e M M Senovilla$^2$ }

\address{$^1$Department of Mathematics, University of Miami, Coral Gables, FL, U.S.A.}
\eads{\mailto{galloway@math.miami.edu}}

\medskip
\address{$^2$ F\'{\i}sica Te\'orica, Universidad del Pa\'{\i}s Vasco, Apartado 644, 48080 Bilbao, Spain}
\eads{\mailto{josemm.senovilla@ehu.es}}


\begin{abstract}
Standard singularity theorems are proven in Lorentzian manifolds of arbitrary dimension $n$ if they contain closed trapped submanifolds of {\em arbitrary} co-dimension. By using the mean curvature vector to characterize trapped submanifolds, a unification of the several possibilities for the boundary conditions in the traditional theorems and their generalization to arbitrary co-dimension is achieved. The classical convergence conditions must be replaced by a condition on sectional curvatures, or tidal forces, which reduces to the former in the cases of co-dimension 1, 2 or $n$.
\end{abstract}

\pacs{04.20.Cv, 04.20.Dw, 04.50.-h, 02.40.Ky}

\section{Introduction}
The celebrated Hawking-Penrose singularity theorem \cite{HP}, see also \cite{BEE,HE,Kr,O,S,W}, proves causal geodesic incompleteness of 4-dimensional spacetimes under strong-energy, causality and generic conditions if there is one of the following:
\begin{itemize}
\item a closed achronal set without edge,
\item a closed trapped surface,
\item a point with re-converging light cone.
\end{itemize}
It is interesting to note that the first case corresponds to a hypersurface (co-dimension 1 submanifold), the second case to a surface (co-dimension 2 submanifold) and the last case to a point (co-dimension $4$, in 4-dimensional spacetimes). 
The co-dimension 3 case ---a closed spacelike curve---, however, is missing in this list. One can wonder why. 

In this communication we will show that the missing case can be included by using the geometrical characterization of trapped submanifolds provided by the mean curvature vector, and thereby one can further prove the singularity theorems in spacetimes of arbitrary dimension with closed trapped submanifolds of arbitrary co-dimension. This was conjectured in \cite{S0}, where the main ideas and the strategy to be used were presented.

\section{The mean curvature vector and trapped submanifolds}
To fix nomenclature and notation, let $\espaitemps$ be an $n$-dimensional causally orientable 
manifold with Lorentzian metric $g_{\mu\nu}$ of signature $-,+,+,\dots ,+$ ($\mu,\nu,\dots =1,\dots ,n$).
Let $\zeta$ be a (for simplicity) connected $(n-m)$-dimensional
submanifold with local intrinsic coordinates $\{\lambda^A\}$ ($A,B,\dots = m+1,\dots ,n$) imbedded in $\varietat$ by the smooth parametric equations
$x^{\alpha} =\Phi ^{\alpha}(\lambda^A) $
where $\{x^{\alpha}\}$ are local coordinates for $\varietat$.
The tangent vectors $\vec{e}_A$ of $\zeta$ are locally given by
\bean
\vec{e}_A \equiv e^{\mu}_A \left.\frac{\partial}{\partial x^{\mu}}
\right\vert _{\zeta} \equiv
\frac{\partial \Phi^{\mu}}{\partial \lambda^A}
\left.\frac{\partial}{\partial x^{\mu}}\right\vert_{\zeta}
\eean
and the first fundamental form of $\zeta$ in $\varietat$ is:
$$\gamma _{AB}\equiv \left.g_{\mu \nu}\right\vert_{\zeta}\frac{\partial \Phi^{\mu}}
{\partial \lambda^A}\frac{\partial \Phi^{\nu}}{\partial \lambda^B}.$$
We assume that $\zeta$ is spacelike and therefore $\gamma _{AB}$ is positive definite. 
Any one-form $n_{\mu}$ defined on $\zeta$ and orthogonal to the tangent vectors
($n_{\mu}e^{\mu}_A=0$) is called a normal one-form to $\zeta$. At each point on $\zeta$ there are $m$ linearly independent normal one-forms. If $m>1$ all of these can be chosen to be null if desired.

The orthogonal splitting into directions tangential or normal 
to $\zeta$ leads to the standard formula \cite{Kr,O}:
$$
\nabla_{\vec{e}_{A}}\vec{e}_{B}=\overline{\Gamma}^{C}_{AB}\vec{e}_{C}-\vec{K}_{AB}
$$
where $\overline{\Gamma}^{C}_{AB}$ are the symbols of the Levi-Civita 
connection $\overline\nabla$ of 
$\gamma$
and $\vec{K}_{AB}$ is the shape tensor of $\zeta$ in $\espaitemps$. Observe that $\vec{K}_{AB}=\vec{K}_{BA}$ and it is
orthogonal to $\zeta$. The contraction of $\vec{K}_{AB}$ with any normal one-form
$$
K_{AB}(\vec n)\equiv n_{\mu}K^{\mu}_{AB}= -n_{\mu}e^{\nu}_A\nabla_{\nu}e^{\mu}_B = e^{\mu}_Be^{\nu}_A\nabla_{\nu}n_{\mu} 
$$
is the second fundamental form with respect to $n_{\mu}$ of $\zeta$ in $\espaitemps$.

The mean curvature vector of $\zeta$ in $\espaitemps$ \cite{O,Kr} is the trace of the shape tensor
$$
\vec{H}\equiv \gamma^{AB}\vec{K}_{AB}
$$
where $\gamma^{AB}$ is the contravariant metric on $\zeta$: $\gamma^{AC}\gamma_{CB}=\delta^{A}_{B}$. By definition, $\vec H$ is orthogonal to $\zeta$. Its contraction with any normal one-form
$$
\theta (\vec n) \equiv n_{\mu}H^\mu =\gamma^{AB}K_{AB}(\vec n)
$$
is called the expansion of $\zeta$ along $\vec n$, and is the trace of the corresponding second fundamental form.

A spacelike submanifold $\zeta$ is said to be {\em future trapped} (f-trapped from now on) if $\vec H$ is timelike and future-pointing everywhere on $\zeta$, and similarly for past 
trapped. This is equivalent to the condition that the expansions relative to {\em all} possible future-pointing normal one-forms are negative. Observe that the extreme case where $n=m$ ($\zeta$ is a point) can be somehow considered to be included here if the expansion along every possible null geodesic emanating from $\zeta$ becomes negative ---as required in the Hawking-Penrose singularity theorem. If $\vec H$ is future-pointing causal everywhere on $\zeta$ then $\zeta$ is said to be weakly future-trapped, while if it is null and future-pointing on $\zeta$ then $\zeta$ is called marginally future-trapped. In these cases the expansions are non-positive. The extreme case with $\vec H=\vec 0$ on $\zeta$ corresponds to a mininal submanifold. (Weakly, marginally) trapped submanifolds of arbitrary co-dimension $m$ were considered in \cite{MS,S0,S1} and they appear to have many common properties independent of $m$.

\section{Existence of points focal to submanifolds of arbitrary dimension}
In order to prove singularity theorems based on trapped submanifolds of arbitrary co-dimension the only intermediate result that needs to be generalized is the existence of focal points along normal geodesics. This will bring in the restrictions on the curvature that produce the necessary convergence of neighbouring geodesics. We will concentrate on future-trapped submanifolds, but the past case can be considered similarly.

Let $n_{\mu}$ be any {\em future-pointing} normal to the spacelike submanifold $\zeta$,
normalized such that $n_{\mu}u^{\mu}=-1$ for a fixed timelike unit vector field $\vec u$,
and let $\gamma$ denote a geodesic curve tangent to $n^\mu$ at $\zeta$. Let $u$ be the affine parameter along $\gamma$, and set $u=0$ at $\zeta$. We denote by $N^{\mu}$ the geodesic vector field tangent to $\gamma$ and by $\vec E_{A}$ the vector fields defined by parallelly propagating $\vec{e}_{A}$ along $\gamma$. Observe that $\vec{E}_{A}|_{u=0}=\vec{e}_{A}$, $N_{\mu}|_{u=0}=n_{\mu}$, and that $N_{\mu}E^\mu_{A}=0$ for all $u$. By construction $g_{\mu\nu}E^\mu_{A}E^\nu_{B}$ is independent of $u$, so that $g_{\mu\nu}E^\mu_{A}E^\nu_{B}=g_{\mu\nu}e^\mu_{A}e^\nu_{B}=\gamma_{AB}$ and one can define along $\gamma$
$$
P^{\nu\sigma}\equiv \gamma^{AB}E^\nu_{A}E^\sigma_{B}
$$
such that $P^{\nu\sigma}=P^{\sigma\nu}$ and $P^{\mu}{}_{\mu}=n-m$. At $u=0$ this is the projector to $\zeta$.

\begin{prop}\label{focal}
Let $\zeta$ be a spacelike submanifold of co-dimension $m$, and let $n_{\mu}$ be a future-pointing normal to $\zeta$. If $\theta (\vec n)\equiv (m-n) c <0$ and the curvature tensor satisfies the inequality
\be
R_{\mu\nu\rho\sigma}N^\mu N^\rho P^{\nu\sigma}\geq 0 \label{cond}
\ee
along $\gamma$, then there is a point focal to $\zeta$ along $\gamma$ at or before $\gamma|_{u=1/c}$, provided $\gamma$ is defined up to that point. 
\end{prop}
\proof
As null geodesics are also considered here, the energy index form (or Hessian) \cite{Kr,O} ---rather than the length index form--- is appropriate. The energy index form of a geodesic $\gamma$ orthogonal to $\zeta$ is the symmetric bilinear form $I(\cdot,\cdot)$ acting on vector fields that vanish at (say) $u=b$ and are orthogonal to $\gamma$ defined by
\be\label{index}
\hspace*{-.7in}I(\vec V,\vec W) \equiv \int_{0}^b\left[(N^\rho \nabla_{\rho}V^\mu )(N^\sigma\nabla_{\sigma}W_{\mu})- N_{\rho}R^\rho{}_{\mu\nu\tau}V^\mu N^\nu W^\tau \right]du
+K_{AB}(\vec n)v^Aw^B
\ee
where $\vec v =\vec V|_{u=0}$, $v_{A}=v_{\mu}e^\mu_{A}$ is the part of $\vec v$ tangent to $\zeta$, and analogously for $\vec W$ and $\vec w$. 

Let $\vec{X}_{A}\equiv (1- c u) \vec{E}_{A}$, that is $\vec{X}_{A}$ are vector fields orthogonal to $\gamma$ such that $\vec{X}_{A}|_{u=0}=\vec{e}_{A}$ and $\vec{X}_{A}|_{u=1/c}=\vec 0$. Then
$$
I(\vec{X}_{A},\vec{X}_{B})=\int_{0}^{1/c} \left[c^2\gamma_{AB} - (1-c u)^2 N_{\rho}R^\rho{}_{\mu\nu\tau}E^\mu_{A} N^\nu E^\tau_{B} \right]du + K_{AB}(\vec n)
$$
and using $\gamma^{AB}K_{AB}(\vec n)=\theta(\vec n)=(m-n)c$ one easily gets
\be
\gamma^{AB}I(\vec{X}_{A},\vec{X}_{B})=-\int_{0}^{1/c} (1-c u)^2N_{\rho}R^\rho{}_{\mu\nu\tau} N^\nu P^{\mu\tau} du \, .\label{average}
\ee
Condition (\ref{cond}) implies that this is non-positive. However, standard results \cite{Kr,O} state that there is no focal point along $\gamma$ for $u\in (0,1/c]$ if and only if the energy index form $I(\cdot,\cdot)$ is positive semi-definite, with $I(\vec V,\vec V)=0$ only if $\vec V$ is proportional to $\vec N$ on $\gamma$. Given that $\gamma_{AB}$ is positive definite it follows from $\gamma^{AB}I(\vec{X}_{A},\vec{X}_{B})\leq 0$ that the energy index form cannot have such a property, and therefore there exists a point focal to $\zeta$ along $\gamma$ for $u\in (0,1/c]$.  \fin

{\bf Remarks:} 
\begin{enumerate}
\item Notice that for spacelike hypersurfaces, when the co-dimension $m=1$, there is a unique timelike orthogonal direction $n_{\mu}$. Then $P_{\mu\nu}=g_{\mu\nu}-(N_{\rho}N^\rho)^{-1}N_{\mu}N_{\nu}$ and the condition (\ref{cond}) reduces to simply
$R_{\mu\nu}N^\mu N^\nu \geq 0$, that is to say, the {\em timelike convergence condition} \cite{BEE,HE,Kr,O,S,W} along $\gamma$.
\item For co-dimension $m=2$, there are two independent null normal directions at each point of $\zeta$, say $n_{\mu}$ and $\ell_{\mu}$. By propagating $\ell_{\mu}$ along $\gamma$ one can define the null vector field $L_{\mu}$ on $\gamma$. Then, $P_{\mu\nu}=g_{\mu\nu}-(N_{\rho}L^\rho)^{-1}(N_{\mu}L_{\nu}+N_{\nu}L_{\mu})$ and again the condition (\ref{cond}) reduces to simply $R_{\mu\nu}N^\mu N^\nu \geq 0$, that is to say, the {\em null convergence condition} \cite{BEE,HE,Kr,O,S,W} along $\gamma$.
\item For co-dimension $m>2$, the interpretation of condition (\ref{cond}) can be given physically in terms of tidal forces, or geometrically in terms of sectional curvatures \cite{Ei,Kr,O}. A completely analogous condition was used in \cite{G} for the proper Riemannian case in connection with minimal submanifolds (of arbitrary dimension). 

For a timelike {\em unit} normal $n_{\mu}$ one has
$$
R_{\mu\nu\rho\sigma}n^\mu e^\nu n^\rho e^\sigma =  k(n,e)   (n_{\rho}n^\rho)(e_{\rho}e^\rho) =
 - k(n,e)  (e_{\rho}e^\rho)
$$
where $n_{\mu}e^\mu =0$, and $k(n,e)$ is called the sectional curvature relative to the plane spanned by $\vec n$ and $\vec e$. Therefore, by choosing $n-m$ mutually orthogonal vectors $\vec e_{A}$ tangent to $\zeta$ condition (\ref{cond}) states that {\em the sum of the $n-m$ sectional curvatures relative to a set of independent and mutually orthogonal timelike planes aligned with $n_{\mu}$ is non-positive, and remains so along $\gamma$}. In physical terms, this is a statement about the attractiveness of the gravitational field, on average. The tidal force ---or geodesic deviation--- in directions initially tangent to $\zeta$ is attractive on average, in such a way that the overall result is a tendency to converge.

For a null normal $n_{\mu}$ one may consider analogously,  
$$
R_{\mu\nu\rho\sigma}n^\mu e^\nu n^\rho e^\sigma =  k(n,e)(e_{\rho}e^\rho)
$$
where $n_{\mu}e^\mu =0$, and $k(n,e)$ is called the {\em null} sectional curvature relative to the plane spanned by $\vec n$ and $\vec e$.  Note, as per the standard definition in this case \cite{BEE},
the minus sign is omitted.  Thus, the interpretation above remains essentially the same (but with ``non-positive" now replaced by ``non-negative") by considering both the sectional curvatures and planes to be null.
\end{enumerate}

The curvature condition (\ref{cond}) in Proposition \ref{focal} can in fact be weakened.    
It is sufficient that it hold on the average in a certain sense. Only a milder, integrated version, is needed, as we now briefly discuss.
Thus, the condition that tidal forces should be attractive along $\gamma$ can be substantially relaxed.

For the following, let the notation be as in Proposition \ref{focal}.

\begin{prop}\label{focal2}
Let $\zeta$ be a spacelike submanifold of co-dimension $m$, and let $n_{\mu}$ be a future-pointing normal to $\zeta$.  If, along $\g$ (assumed to be future complete) the curvature tensor satisfies,
\be\label{cond2}
\int_0^{\infty} R_{\mu\nu\rho\sigma}N^\mu N^\rho P^{\nu\sigma} du >  \theta (\vec n) \,,
\ee
then there is a point focal to $\zeta$ along 
$\gamma$. 
\end{prop}

Observe that there is no restriction on the sign of $\theta (\vec n)$.
We also note that, unlike Proposition \ref{focal}, this proposition does not restrict the location of the focal point, but this will not be needed to prove singularity theorems under weaker curvature conditions (such as (\ref{cond3}) below.) 

\proof Let $f = f(u), u \ge 0$ be the solution to the initial value problem,
\bean
&(n-m) f'' + r(u)f = 0 \,, \\
&f(0)  = 1, \quad f'(0) = \frac{\theta (\vec n)}{n-m}  \,,
\eean
where $r \equiv R_{\mu\nu\rho\sigma}N^\mu N^\rho P^{\nu\sigma}$.  Given that (\ref{cond2}) holds, it follows immediately from Lemma~3 in \cite{G} that $f$ has a zero on $(0,\infty)$, i.e., there exists
$b > 0$ such that $f(b) = 0$.

Now let  $\vec{X}_{A}\equiv  f \vec{E}_{A}$ on $[0,b]$.  Substituting into (\ref{index}), gives,
$$
I(\vec{X}_{A},\vec{X}_{B})=\int_{0}^{b} \left[f'^2 \gamma_{AB} - f^2 N_{\rho}R^\rho{}_{\mu\nu\tau}E^\mu_{A} N^\nu E^\tau_{B} \right]du + K_{AB}(\vec n)  \,.
$$
Contracting with $\g^{AB}$ and using $f'^2 = (ff')' - f f''$, we obtain,
$$
\gamma^{AB}I(\vec{X}_{A},\vec{X}_{B})= -\int_{0}^{b}[(n-m)f'' + r(u)f] f du 
+(n-m) f f' |_0^{b} + \theta (\vec n) = 0  \,.
$$
The remainder of the argument is just as in the proof of Proposition \ref{focal}.~\fin

\section{Main results: singularity theorems}
We start by proving the generalization of the Penrose singularity theorem \cite{P}, see also \cite{BEE,HE,Kr,O,S,W}, which is the first of the ``modern'' theorems. Then, we will also prove the generalization of  the more elaborated Hawking-Penrose theorem.

Recall that for any set $\zeta$, $E^+(\zeta)\equiv J^+(\zeta)\backslash I^+(\zeta)$, using the standard notation for the causal $J^+(\zeta)$ and chronological $I^+(\zeta)$ futures of $\zeta$, see e.g. \cite{BEE,HP,HE,Kr,P1,S}.
\begin{prop} 
Let $\zeta$ be a closed f-trapped submanifold of co-dimension $m>1$, and assume that the curvature tensor satisfies the inequality
$$
R_{\mu\nu\rho\sigma}N^\mu N^\rho P^{\nu\sigma}\geq 0
$$
for any future-pointing null normal one-form $n_{\mu}$. Then, either $E^+(\zeta)$ is compact, or the spacetime is future null geodesically incomplete, or both.\label{2}
\end{prop}
{\bf Remark:} The case with $m=1$ is not included here because it is trivial. If $\zeta$ is a spacelike hypersurface, then $E^+(\zeta)\subset \zeta$ ---and actually $E^+(\zeta)= \zeta$ if $\zeta$ is achronal---, and the compactness of $E^+(\zeta)$ follows readily without any further assumptions.

\proof Assume that $\espaitemps$ is future null geodesically complete. As $\zeta$ is f-trapped one has $\theta(\vec n)=(m-n)c<0$ for any future-pointing null normal one-form $n_{\mu}$. Let $(m-n)C$ be the maximum value of all possible $\theta (\vec n)$ on the compact $\zeta$. Due to Proposition \ref{focal} every null geodesic emanating orthogonally from $\zeta$ will have a focal point at or before the affine parameter reaches the value $1/C$. Standard results \cite{BEE,HE,Kr,O,S,W} imply that these null geodesics enter $I^+(\zeta)$ from the focal point on. Let ${\cal K}$ be the set of points reached by all these null geodesics up to the affine parameter $1/C$ inclusive, so that ${\cal K}$ is compact. Obviously $E^+(\zeta)\subset {\cal K}$, and thus it is enough to prove that $E^+(\zeta)$ is closed. Let $\{p_{i}\}$ be an infinite sequence of points $p_{i}\in E^+(\zeta)$ and let $p$ be their accumulation point on the compact ${\cal K}$. $p$ cannot be in $I^+(\zeta)$ as otherwise, 
$I^+(\zeta)$ being open, there would be a neighbourhood of $p$ within $I^+(\zeta)$ containing some of the $p_{i}$, which is impossible as $p_{i}\in E^+(\zeta)$. But $p$ certainly is in ${\cal K}\subset J^+(\zeta)$. Hence, $p\in E^+(\zeta)$ showing that $E^+(\zeta)$ is closed and thus compact.\fin

The analogue of the Penrose singularity theorem can now be proven.
\begin{theorem}\label{genpen}
If $\espaitemps$ contains a non-compact Cauchy hypersurface $\Sigma$ and a closed f-trapped submanifold $\zeta$ of arbitrary co-dimension, and if condition (\ref{cond}) holds along every
future directed null geodesic emanating orthogonally from $\zeta$, then $\espaitemps$ is future null geodesically incomplete.
\end{theorem}
\proof The proof is standard \cite{BEE,HE,O,P,S,W}, so we just sketch it. If $\espaitemps$ were null geodesically complete $E^+(\zeta)$ would be compact due to Proposition \ref{2}. But the spacetime is globally hyperbolic so that 

(i) $E^+(\zeta)=\partial J^+(\zeta)$ is the boundary of the future set $J^+(\zeta)$ and therefore a proper achronal boundary, which are known to be imbedded
submanifolds (without boundary) \cite{P1,HP,HE,Kr,S,W}; and

(ii) the manifold is the product $\varietat = \mathbb{R} \times \Sigma $, see \cite{BS} and references therein.

Then the canonical projection on $\Sigma$ of the compact achronal $E^+(\zeta)$ would have to have a boundary, ergo the contradiction.\fin
\begin{theorem}\label{claim}  The conclusion of Propositon \ref{2},  and hence, of Theorem \ref{genpen},  remains valid if the curvature
condition and the trapping condition assumed there are jointly replaced by
\be\label{cond3}
\int_0^{a} R_{\mu\nu\rho\sigma}N^\mu N^\rho P^{\nu\sigma} du >  \theta (\vec n) \,,
\ee
along each future inextendible null geodesic $\g \!\!: [0,a) \to \varietat$ emanating orthogonally from~$\zeta$ with initial tangent $n^{\mu}$.
\end{theorem}
\proof  We shall assume that $\zeta$ is acausal.  (The case in which $\zeta$ is not acausal can be handled by considering a suitable finite cover of $\zeta$ by acausal subsets.)  To prove the claim,  it is sufficient to show that $E^+(\zeta)$ is compact  
under the assumption of future null geodesic completeness. Under this assumption, Proposition \ref{focal2} implies that there is a focal point along each future directed null geodesic emanating orthogonally from $\zeta$.  Since cut points to $\zeta$ \footnote{See \cite{Ku} for relevant definitions and properties in the co-dimension two case; the higher co-dimension case works similarly.} come at or before focal points,  there is a null cut point along each of these null geodesics.  The compactness of $\zeta$ and the fact that the affine distance to each null cut point is upper semi-continuous, as a function of points in $\zeta$, 
imply that there exists $C^* > 0$ such that each cut point occurs at or before the affine value $C^*$. Let ${\cal K}$ be the set of points reached by all these null geodesics up to the affine parameter $C^*$ inclusive, so that ${\cal K}$ is compact.   Since $E^+(\zeta)\subset {\cal K}$, and, as argued in Proposition \ref{2}, $E^+(\zeta)$ is closed, we have that $E^+(\zeta)$ is compact.~\finn

Thus, for example, even if $\zeta$ is only weakly or marginally f-trapped, or minimal,
the conclusion  of future null geodesic incompleteness in Theorem \ref{genpen} still holds, provided the inequality (\ref{cond}) is strict at least at 
one point on each future directed null geodesic $\g$ emanating orthogonally from $\zeta$.

Let us finally consider the more powerful Hawking-Penrose singularity theorem \cite{HP}. As is known, this theorem is based on the Hawking-Penrose Lemma, proven in \cite{HP}, see also \cite{HE,Kr,BEE,S}, stating that the following three statements cannot hold simultaneously in a given spacetime
\begin{itemize}
\item every endless causal geodesic has a pair of conjugate points,
\item there are no closed timelike curves (chronology condition),
\item there is an achronal set $\eta$ such that $E^+(\eta)$ is compact.
\end{itemize}
This result holds independently of the dimension, and the achronal set $\eta$ can also be anything as long as $E^+(\eta)$ is compact. Thus, one only has to ensure that the standard result leading to the compactness of $E^+(\eta)$ is achieved if there is a f-trapped submanifold of arbitrary dimension. This is proven now ---for the definition of strong causality refer to any standard reference \cite{BEE,HE,Kr,O,P1,S,W}---.
\begin{prop}
If $\espaitemps$ is strongly causal and there is a closed f-trapped submanifold $\zeta$ of arbitrary co-dimension $m>1$ such that condition (\ref{cond}) holds along every null geodesic emanating orthogonally from $\zeta$, then either $E^+(E^+(\zeta)\cap \zeta)$ is compact, or the spacetime is null geodesically incomplete, or both.
\label{3}
\end{prop}
{\bf Remark:} Thus, $\eta \equiv E^+(\zeta)\cap \zeta$ provides the needed set in the Hawking-Penrose Lemma above, the achronality of $\eta$ being a direct consequence of the achronality of $E^+(\zeta)$.
\proof Assume that $\espaitemps$ is null geodesically complete, hence $E^+(\zeta)$ is compact due to Proposition \ref{2} so that $\eta$ is also compact. The set $\eta$ is also non-empty, as otherwise $\zeta\subset I^+(\zeta)$ against the assumption of strong causality \cite{BEE,HE,S}. One only needs to prove that actually $E^+(\eta)=E^+(\zeta)$, ergo compact. This can be done in standard fashion \cite{BEE,S} by covering the compact $\zeta$ with convex normal neighbourhoods such that the piece of $\zeta$ on each of them is achronal, extracting a finite sub-cover, and then showing that $I^+(\zeta)\subset I^+(\eta)$, leading immediately to $I^+(\zeta)= I^+(\eta)$. Consider now $q\in J^+(\zeta)$. From the previous equality, if $q\in I^+(\zeta)$ then $q\in I^+(\eta)\subset J^+(\eta)$. If $q\notin I^+(\zeta)=I^+(\eta)$, thene there is a point $p\in \zeta$ with $q\in E^+(p)$. It is obvious that $p\notin I^+(\zeta)$ because $q\notin I^+(\zeta)$, hence $p\in \zeta-I^+(\zeta)$ ergo $p \in E^+(\zeta)\cap \zeta =\eta$ so that $q\in J^+ (\eta)$, implying that $J^+(\zeta)=J^+(\eta)$. Finally, $E^+(\eta)=J^+(\eta)-I^+(\eta)=J^+(\zeta)-I^+(\zeta)=E^+(\zeta)$ as required.~\fin

From this result the analogue of the Hawking-Penrose theorem follows at once.
\begin{theorem}
If the chronology, generic and strong energy conditions hold and there is a closed f-trapped submanifold $\zeta$ of arbitrary co-dimension such that condition (\ref{cond}) holds along every null geodesic emanating orhogonally from $\zeta$ then the spacetime is causal geodesically incomplete.\finn
\end{theorem}
{\bf Remark:} Of course, for co-dimension $m=1$ there are no null geodesics orthogonal to $\zeta$ and there is no need to assume (\ref{cond}) nor anything concerning the mean curvature vector of $\zeta$, as in Remark (i) to Proposition \ref{2}. For co-dimension $m=2$ the condition (\ref{cond}) is actually included in the strong energy condition as explained in Remark (ii) to Proposition \ref{focal}. The same happens for co-dimension $m=n$. These three cases cover the original Hawking-Penrose theorem.

In principle, any other singularity theorem assuming a closed trapped surface can be appropriately generalized to arbitrary co-dimension by means of Propositions \ref{focal}--\ref{3}.

\section{Discussion with some applications}
The main application of these theorems is, of course, to higher dimensional spacetimes, and therefore they can be used in the fashionable Kaluza-Klein/string/supergravity/M-type theories. In dimension 11, say, there are now 10 different possibilities for the boundary condition in the theorems, in contrast with the classical three possibilities. All this should be explored and can have relevance in connection with the {\em compactified} extra-dimensions.

For example, the singularity theorems proven in this paper reinforce the arguments put forward by Penrose \cite{P2} about the {\em classical} instability of spatial extra-dimensions, as they might develop singularities within a tiny fraction of a second. In a 10-dimensional spacetime, the argument in \cite{P2} needs some ad-hoc splittings, and some restrictions on the Ricci tensor, that can be avoided by using the Theorems proven herein. It is enough that the compact extra-dimensional space, or {\em any} of its compact less-dimensional subsets, satisfy the trapping condition, and the restriction on Ricci curvatures can be replaced by the appropriate (averaged) condition on tidal forces. All in all, the basic argument of Penrose acquires a wider applicability and requires less restrictions by using the singularities proven to develop under the existence of compact submanifolds of any dimension.

Even in the traditional 4-dimensional Lorentzian manifolds of any classical gravitational theory, the new theorem may have some applications when considering the case of a closed trapped {\em curve}. Observe that these are just curves whose acceleration vector is timelike. An obvious relevant example, as explained in \cite{S0}, is the case of spacetimes with whole cylindrical symmetry \cite{Exact}, 
expressed in local coordinates by the line-element
\be
ds^2=-A^2dt^2+B^2d\rho^2+F^2d\varphi^2+E^2dz^2, \label{cyl}
\ee
where $\partial_{\varphi},\partial_{z}$ are spacelike commuting Killing 
vectors.
The coordinate $\varphi$ is 
closed with standard periodicity $2\pi$. The cylinders given by constant 
values of $t$ and $\rho$ are geometrically preferred 2-surfaces, however, these cylinders are {\em not} compact in general, so that they have no {\em direct} 
implication in the development of geodesic incompleteness. 

Nevertheless, the spacelike curves defined by constant values of 
$t,\rho$ and $z$ are also geometrically distinguished and certainly {\em closed}. Their mean curvature vector is easily seen to be proportional to $dF$. Thus, the 
causal character of the gradient of {\em only} the function $F$, which is the norm of the circular Killing vector $\partial_{\varphi}$, characterizes the trapping of these closed circles.  Thereby, many results on incompleteness of geodesics can be found. Moreover, there arises a new hypersurface, defined as the set of points where $dF$ is null, which is a new type of horizon, being a boundary separating the trapped from the untrapped circles, and containing marginally trapped circles.

We now consider an application of Theorem \ref{genpen} (or, more precisely, its time-dual) to the existence of initial singularities in asymptotically de Sitter spacetimes with compact Cauchy hypersurfaces.
This, as we shall see, also involves trapped circles. 
In \cite{ AG, G2}, results were obtained that establish a connection between Cauchy hypersurfaces with nontrivial topology and the occurrence of past singularities in such spacetimes. This behavior is well illustrated by dust filled FLRW solutions to the Einstein equations \cite{Exact} with positive cosmological constant, $\Lambda >0$; cf., the discussion \cite{G2}.   Such models with spherical spatial topology (de Sitter
spacetime being the limiting case) need not have past singularities, whereas models which have, e.g.,
toroidal spatial topology, do  have past singularities.   We now present another general result along these lines.

We shall say that a group $G$ is {\it sufficiently large} provided it has a nontrivial normal subgroup
$H$ such that $|G/H| = \infty$ (i.e., the cardinality of $G/H$ is not finite.)  The fundamental group $\pi_1(T^k) = \mathbb{Z}^k$ of the $k$-torus $T^k$, $k \ge 2$, is an example of a sufficiently large group. 
 
\begin{theorem}  Let  $\espaitemps$ have dimension $n \ge 3$,
with all null sectional curvatures non-negative. 
Suppose $\Sigma$ is a compact Cauchy hypersurface for $\espaitemps$  which is expanding to the future in all directions, i.e., which has positive definite second fundamental form with respect to the future
pointing normal.  Then, if $\pi_1(\Sigma)$ is sufficiently large, $\espaitemps$ is past null geodesically incomplete.
\end{theorem}

\proof  Let $H$ be a nontrivial normal subgroup of $\pi_1(\Sigma)$ such that $|\pi_1(\Sigma)/H| = \infty$.
Since $\Sigma$ is compact we can minimize arc length in the free homotopy class of a nontrival
element of $H$ to obtain a closed geodesic $\s$ in $\Sigma$.  Since $\Sigma$ has negative definite second fundamental form with respect to the {\it past} pointing unit normal, one easily verifies that $\s$
is a past-trapped co-dimension $n-1$ submanifold in $\espaitemps$.  If $\espaitemps$ were past null geodesically complete then the time-dual of Proposition \ref{2} and the curvature condition would
imply that $E^-(\s)$ is compact.  Since all the Cauchy hypersurfaces of $\espaitemps$ are compact \cite{BS}, this does not
directly lead to a contradiction.  But now we pass to a covering spacetime. 

By standard covering space theory  there exists a covering manifold $\tilde \Sigma$ of $\Sigma$, with covering map $p:\tilde \Sigma \to \Sigma$, such that the induced map on fundamental groups 
$p_*: \pi_1(\tilde \Sigma) \to H \subset \pi_1(\Sigma)$ is an isomorphism. Since $|\pi_1(\Sigma)/H| = \infty$, this is an infinite sheeted covering, and hence $\tilde \Sigma$ is non-compact.  We know by the global hyperbolicity
of $\espaitemps$ that $\varietat$ is diffeomorphic to $\mathbb{R} \times \Sigma$ \cite{BS}, and hence that the fundamental groups of $\Sigma$ and $\varietat$ are isomorphic.  From this it follows that there is a covering spacetime $(\tilde\varietat, \tilde g)$, with covering map $P:\tilde\varietat \to \varietat$, such that (i) $P$ is a local isometry, ($\tilde g$ is the pullback of $g$ via $P$), and (ii) $\tilde\varietat$ contains $\tilde \Sigma$ as a Cauchy hypersurface, such that $P|_{\tilde  \Sigma} = p$.   Now, since $p_*$ is an isomorphism, $\s$ lifts to a closed geodesic $\tilde \s$ in $\tilde \Sigma$, which, because $P$ is a local
isometry, is past trapped in  $(\tilde\varietat, \tilde g)$.  Then, since the curvature condition lifts
to $(\tilde\varietat, \tilde g)$, the time-dual of Theorem \ref{genpen} implies that there exists a past incomplete null geodesic in $(\tilde\varietat, \tilde g)$.  This projects, via $P$, to a past incomplete null geodesic in $\espaitemps$.~\fin

Of course, this theorem has a dual version to the future, if the compact Cauchy hypersurface is contracting.

\section*{Acknowledgements}
Comments from M Mars and M S\'anchez are gratefully acknowledged. Supported by grants FIS2004-01626 (MICINN), GIU06/37 (UPV/EHU), and 
DMS-0708048 (NSF).

\end{document}